\documentclass[pre,twocolumn,aps,groupedaddress]{revtex4-2}
\usepackage[T1]{fontenc}
\usepackage[latin9]{inputenc}
\usepackage{array}
\usepackage{booktabs}
\usepackage{mathrsfs}
\usepackage{mathbbol}
\usepackage{multirow}
\usepackage{amsmath}
\usepackage{amsthm}
\usepackage{amssymb}
\usepackage{stmaryrd}
\usepackage{graphicx}
\usepackage{latexsym}
\usepackage{textcomp}
\usepackage{mathtools}
\usepackage{xcolor}
\usepackage[allcolors=blue,colorlinks=true]{hyperref}
\usepackage{lineno}
\usepackage{stackrel}
\usepackage[toc,page,titletoc]{appendix}
\usepackage{bm}
\definecolor{mycolor1}{rgb}{0.1, 0.6, 0.6}

\begin{document}

\title{Optimum dissipation by cruising in dense suspensions}

\author{Pappu Acharya}
\email{pappu.acharya@teokem.lu.se}

\affiliation{Theoretical Chemistry, Lund University, Lund SE-221 00, Sweden}

\author{Martin Trulsson}
\email{martin.trulsson@teokem.lu.se}

\affiliation{Theoretical Chemistry, Lund University, Lund SE-221 00, Sweden}

\begin{abstract}
 Dense suspensions tend to shear jam at large packing fractions. However, it has recently been shown that various oscillation 
protocols can unjam as well as reduce viscosity and dissipation. In this numerical 
work, we ask ourselves what is the optimum shear protocol in terms of dissipation. 
We show that many cruising protocols' dissipation are similar to shear protocols with a steady primary shear and superimposed cross oscillations, even though the latter's viscosity reduction is more considerable. Furthermore, we find that alternating between primary and perpendicular oscillations yields a much higher dissipation than the two protocols mentioned above, yet has similar viscosity as the cross-oscillatory one. While self-organization has been shown to minimize viscosity, our findings
challenge the idea that random organization is the underlying mechanism for reducing dissipation. Instead, shear ``fragility'' combined with geometry seems to be the key ingredients,
which explains the counter-intuitive decoupling of the minima of viscosity and dissipation for the cruising protocol. This work paves the way for a new class of highly-energy efficient flow protocols.  
\end{abstract}


\maketitle

\section{Introduction}
Dense suspensions are widespread in our society, both as industrial and consumer products. Studying them is, hence, relevant for a better society.  For example, medical pills formation, vaccines, and manufacturing of ceramic and paste all rely on processes that involve dense suspensions. Dense suspensions are also widespread in nature, from flowing magma to beach sand and mud. Mixing cornstarch and water, creating oobleck, is also an ideal and popular toy experiment to increase interest and educate our kids in natural science.

Dense suspensions are composed of solid, Brownian, or non-Brownian, particles suspended in a fluid with a high volume fraction of the solid phase. As the fraction of solid particles increases, so does also the flow resistance, \emph{i.e.,~}the viscosity. Above a certain packing fraction $\phi_J$, a suspension generally jams and stops to flow \cite{Guezzelli2018review}. The viscosity usually shows an inverse power law divergence as a function of the distance to this jamming point, indicating a second-order non-equilibrium phase transition with an associated diverging cooperative length \cite{Heussinger2010diverg,Trulsson2017diverg,Olsson2020diverg}. Dense suspensions also often show non-Newtonian behavior, where the flow resistance is nonlinear as a function of the flow rate. A sub-linear relationship is denoted as shear-thinning and super-linear shear-thickening. A typical example of a shear-thinning suspension is blood and a shear-thickening is oobleck. 
The latter can even show discontinuous shear-thickening (denoted DST) \cite{Seto2013DST,Jamali2019DST, Dong2017DST,Singh2020DST} where above a certain
shear stress or flow rate a suspension does an abrupt transition from a low-viscous fluid to a high viscous one or even become an elastic solid.
There are advantages and disadvantages of such a DST scenario. An application making use of one of these advantages is Kevlar soaked with such a DST suspension to generate a flexible yet impact-proof space suit, called an STF-Armor \cite{cwalina2015mmod}.
The same concept has also been used to design bullet-proof jackets \cite{wagner2002advanced}.
Despite the attractive fringe benefit of shear thickening, the main drawback comes when a suspension needs to flow. \\

For all dense suspensions, DST or not, it is beneficial if we can alter and facilitate flowability. In the literature, many works have been focused on the rheological properties of dense suspensions and some are able to reduce a suspension's viscosity by more than 90 percent. The procedure to reduce viscosity involves many strategies, \emph{e.g.}~ adding polydispersity \cite{pednekar2018bidisperse} to the particle sizes, changing their shapes \cite{Nagy2017shape,trulsson2018ellip}, adding some lubricating particles, by adding another fluid \cite{xu2013decrease} \emph{etc.}. One recent and successful approach is to
apply a cross shear to the suspension as it does not involve a change of a suspension's composition.  However, there is an
extra energy cost involved in doing so. While many of the previous
works \cite{lin2016tunable,dong2020transition,sehgal2019using} have been focused on reducing the viscosity, only a few focused on the energy needed to propagate the suspension, \emph{i.e.,~}the energy dissipation, \cite{dong2020transition, ness2018shaken}.

In this work, we design and study a large set of numerical shear protocols to find optimal ways of propagating a suspension in terms of energy dissipation (per strain and time). We argue that oscillations are not necessary for the dissipation reduction, rather, a change in the direction of the fluid flow is sufficient, related to the fragility of dense suspensions \cite{Cates1998fragile,seto19fragile,blanc23fragile}. We illustrate this by applying a cruising protocol, which gives a comparable dissipation reduction compared to previously best-reported protocols. This cruising protocol uses only two directions, compared to three for oscillating protocols, and does not involve a flow reversal \cite{Blanc2011reverse,Peters2016reverse}, which are known to reduce the viscosity, but rather shear rotations. Furthermore, we show that the optimal angle for our cruising protocol, \emph{i.e.,~}shear rotations, depends on the packing fraction, where in the vicinity of the steady-state jamming point obtuse angles $(>90^o)$ are superior. We give a clear explanation of why some protocols are better than others with semi-theoretical expressions of the dissipation for each protocol.





\label{section_simulation}
\begin{figure}[t!]
\includegraphics[width=0.9\linewidth]{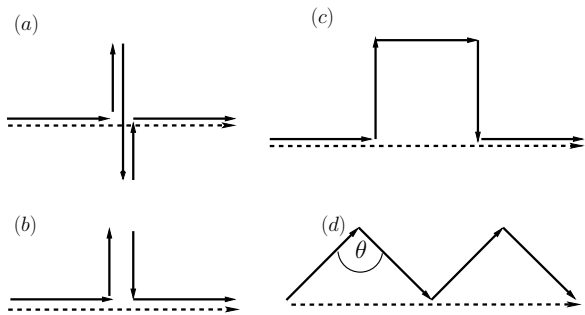}
\caption{Schematic diagram of all four protocols (a) AO or alternating oscillatory cross shear \cite{ness2018shaken}, (b) HO or half oscillatory cross shear, (c) SW or square wave shear and (d) CR or cruising. In all four protocols, we used $xz$ as the primary shear plane while $xy$ as the cross shear plane.}
\label{fig_sch}
\end{figure}
\section{the AO protocol's dissipation}
\color{black} 
We start by re-analyzing one of the previously proposed protocols, denoted AO (Alternating Oscillatory cross shear) \cite{ness2018shaken}.
In this protocol, the system is sheared linearly in the primary direction while an oscillatory shear is applied along the perpendicular direction. A full cycle is divided into two parts where $\alpha$ is related to the time spent in the primary and oscillatory directions. A strain rate of $\dot{\gamma}_{\rm pri}=\dot{\gamma}/(1-\alpha)$ is applied in the primary direction for time $(1-\alpha)T$, yielding a strain in the primary direction equal to $\dot \gamma_{\rm pri} (1-\alpha) T = \dot \gamma T=\gamma_{\rm pri}$. Here, $\dot \gamma$ is the corresponding shear rate for the steady-shear state, yielding the same strain if sheared for a time $T$. During shearing in the primary direction, there is no oscillatory shear along the perpendicular direction.
Then, for a time $\alpha T$, an oscillatory shear with a time period of $\alpha T$ is applied as $\gamma_{\rm cross} = \gamma_{\rm osc} \sin\omega t$, where $\omega=2\pi/\alpha T$ and $t$ is time. We find the equation for dissipation per strain (and time) with $\alpha$ (see the Appendix for details) as 
\begin{equation}
W = \frac{W_{\parallel}+W_{\perp}}{W_{\rm ss}}=\frac{\eta_\parallel}{\eta_{\rm ss}}\left(\frac{1}{1-\alpha}\right)+\frac{\eta_\perp}{\eta_{ss}}\frac{2\pi^2}{f^2}\left(\frac{1}{\alpha}\right),
\label{eq_dissipation_AO_sin_main}
\end{equation} 
where $W_{\parallel}$, $W_{\perp}$, $W_{\rm SS}$, are the energy of dissipation per strain for the primary shear, cross shear, and steady-state (SS) respectively. Here, $\eta_{ss}$ is the steady state viscosity which only depends upon the packing fraction of the system, $f=\gamma_{\rm pri}/\gamma_{\rm osc}$ is the ratio of strain along the primary and perpendicular directions, while, $\eta_\parallel$ and $\eta_\perp$ are the viscosities along the primary and perpendicular directions and can be computed directly from the simulations. 
We plot the dissipation from the simulations as a function of $\alpha$ along with this expression in Fig. \ref{fig_dissipation_AO_appendix}(a) in the Appendix, which shows a good agreement for three different $f$.
As we apply primary shear to the suspension along the flow direction of the $xz$ plane, we denote the primary direction of flow as $xz$ and the oscillatory direction as $xy$ throughout this paper, therefore $\eta_\parallel$ and $\eta_\perp$ can be defined as 
\begin{eqnarray}
\eta_\parallel &=& \frac{\sigma_{xz}}{\dot{\gamma}_{\rm pri}},\\
\nonumber
\eta_\perp &=& \frac{ \int_{0}^\frac{2\pi}{\omega}\sigma_{xy} \cos(\omega t) dt}{\dot{\gamma}_0 \int_{0}^\frac{2\pi}{\omega}\cos^2(\omega t) dt},
\end{eqnarray}
where $\sigma_{kl}$ denotes shear stresses. Note that the viscosities are rate-independent in the over-damped limit and hence do not depend on $\alpha$. 
The minimum point for each $f$ can be calculated by taking a derivative of Eq. (\ref{eq_dissipation_AO_sin_main}) with respect to $\alpha$ and equating it to zero. The real root for all the values of $\eta_\parallel$ and $\eta_\perp$ can then be written as 
\begin{equation}
\alpha = \frac{1}{1+\frac{f}{\sqrt{2}\pi}\left(\frac{\eta_\parallel}{\eta_\perp}\right)}.
\end{equation} 
At increasing $f$, $\eta_\parallel$ increases, while $\eta_\perp$ is roughly constant. Hence, we can conclude that the minimum point in the dissipation curve will shift toward $\alpha=0$ with increasing $f$ as displayed in Fig. \ref{fig_dissipation_AO_appendix}(a) in the Appendix. The minimum value of the dissipation for different $f$ can then be written as
\begin{equation}
W_{min}=\frac{(\sqrt{\eta_\parallel}+\frac{\sqrt{2}\pi}{f}\sqrt{\eta_\perp})²}{\eta_{ss}}.
\end{equation} 
One of our first findings is that even though the AO protocol reduces the viscosity in the primary direction, it comes at a cost in dissipation, where the AO dissipate more than doing standard steady-state. This is at odds with the original AO results of Ref.~\cite{ness2018shaken}, which was due to a subtle error (see the reference's erratum). \\

A similar protocol but with a constant shear rate along the perpendicular direction yields the dissipation as (see the Appendix for details)
\begin{equation}
W_{\rm AO} = \frac{W_{\parallel}+W_{\perp}}{W_{\rm ss}}=\frac{\eta_\parallel}{\eta_{\rm ss}}\left(\frac{1}{1-\alpha}\right)+\frac{\eta_\perp}{\eta_{\rm ss}}\frac{16}{f^2}\left(\frac{1}{\alpha}\right).
\label{eq_dissipation_AO_linear_main}
\end{equation} 
Both protocols' viscosities are very close with the latter slightly less dissipative as $2\pi^2 >16$. We again plot the dissipation from the simulations with $\alpha$ along with this expression in  Fig. \ref{fig_dissipation_AO_appendix}(b) in the Appendix, which shows a good agreement for three different $f$. Nevertheless, both protocols dissipate a large amount of energy (more than the steady state) and are not very fruitful in terms of energy reduction.  \\

We continue by re-analyzing also the protocol denoted SO (Simple oscillatory cross shear), where oscillations instead are superimposed on a primary shear direction in a steady manner. In contrast to the AO protocol, we find the same dissipation (data not shown) from our numerical simulations as reported in Ref.~\cite{ness2018shaken}. For expressions of how dissipation depends on viscosities, see our Appendix.
Given that SO and AO give such different dissipation, this naturally leads to the questions of (i) why AO and SO are so different? and (ii) what is the optimum way of propagating a suspension? We focus on relative dissipation, keeping the strain and time compared to the steady-state protocol, {\emph{i.e.,~}at a fixed global shear rate $\dot \gamma = \gamma_{\rm pri}/T$.
This as dissipation per strain $W$ can, trivially, be reduced by allowing for a slower global shear rate (equivalent to longer shear time). 
This can for example be realized as  $W_{SS} = \eta_{\rm ss} \dot \gamma$ and $\eta_{\rm ss}$ is rate-independent in our model. 
This means we trade time for energy. In practice, one wants to keep both at a minimum. Since  $W_X= \eta_X \dot \gamma$ irrespective of the protocol, the relative dissipation (per strain and time) \emph{i.e.}  $W=W_X/W_{SS}$ is actually measuring a relative ``dissipative'' viscosity \cite{dong2020transition}.


\begin{figure}[t!]
\centering
\includegraphics[width=1.1\linewidth]{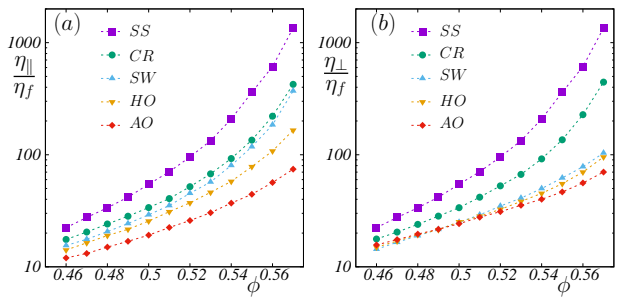}
\caption{Relative viscosity along the (a) parallel and (b) perpendicular directions for different protocols with packing fraction. We also plot the steady state value as a comparison. In all the simulations, we use $\mu = 1$ and $\alpha=0.8,0.7,0.5$ and $0.5$ for AO, HO, SW, and CR respectively. Here the symbols are numerical data while the dashed lines are guides for the eye.
}
\label{fig_viscosity_allprotocol}
\end{figure}
\color{black}
\section{Improving the AO protocol}
The dissipation in the AO protocol is rather high despite the huge viscosity reduction. 
To minimize energy dissipation, we systematically apply different types of protocols to our system and ascertain an optimum protocol. Here we give an introduction to various protocols. For all the protocols, we use strains of $1\%$ in all directions and normalize the dissipation with the propagating strain. We then normalize with the steady-state value at the same global shear rate as discussed earlier.  

\subsection{Half oscillation}
A full oscillation along the perpendicular direction costs an immense amount of energy; therefore, we first try to reduce that by applying half oscillation (see Fig. \ref{fig_sch} (b)). The idea is to reduce the accumulated strain in the cross direction from $4\gamma_{\rm osc}$ to $2\gamma_{\rm osc}$.
In this half oscillation protocol, a full cycle is composed of $1\%$ strain along the primary direction (flow direction of the $xz$ plane), with a shear rate $\dot{\gamma}_{\rm pri} = \dot{\gamma}/(1-\alpha)$ and a time $(1-\alpha)T$. Along the perpendicular direction \emph{i.e.~}$xy$, we shear the system $1\%$ up and $1\%$ down with the applied shear rate $\dot{\gamma}_{\rm osc} = 2 \dot{\gamma}/\alpha$ for a time $\alpha T$. The total time for a cycle is hence $T$ and the strain per cycle in the primary direction is $1\%$ just like in the previously described AO protocol.
For the half oscillation, the relative dissipation relation can be written as (see the Appendix for details)
\begin{equation}
W_{\rm HO}= \frac{W_{\parallel}+W_{\perp}}{W_{\rm ss}}=\frac{\eta_\parallel}{\eta_{\rm ss}}\left(\frac{1}{1-\alpha}\right)+\frac{\eta_\perp}{\eta_{\rm ss}}\frac{4}{f^2}\left(\frac{1}{\alpha}\right).
\label{eq_dissipation_HO_linear_main}
\end{equation} 
Comparing with Eq.~(\ref{eq_dissipation_AO_linear_main}), we see that we only change the prefactor of the second term from $16$ to $4$. However, both $\eta_{\parallel}$ and 
$\eta_{\perp}$ depend on the protocol applied. Similar to the full oscillation, once the two viscosities are known, we can easily minimize the dissipation with respect to $\alpha$.
All reported relative dissipation corresponds to the minimum values. 
The dissipation is slightly less in this case compared to the AO, although the viscosity reduction is lower in the former. We discuss these results later in section \ref{sec_results}.
\subsection{Square wave}
In order to reduce the energy dissipation further, we design a new protocol where the shear path looks like a square wave (SW), as shown in Fig. \ref{fig_sch}(c). This could be considered as a quarter oscillation protocol but differs from the previous two in that it does not include a shear reversal between two primary shears.
In the first cycle, we shear the system along the flow direction of the $xz$ plane in the first half, and then along the flow direction of the $xy$ plane in the second half. In the second cycle, however, we shear the system along the flow direction of the $xz$ plane again in the first half, but in the second half, we shear it along $-xy$. The key idea with this is to further reduce the accumulated strain in the cross direction (per cycle). We discuss this protocol's viscosity and dissipation reductions later in section \ref{sec_results}.
The relative dissipation for a square wave follows a similar trend as for the half oscillation protocol, where we once again affect prefactor of the second term, and can be written as (see the Appendix for details)
\begin{small}
\begin{equation}
    W_{\rm SW} = \frac{\eta_\parallel}{\eta_{\rm ss}}\left(\frac{1}{1-\alpha}\right)+\frac{\eta_\perp}{\eta_{\rm ss}}\left(\frac{1}{\alpha}\right).
    \label{eq_dissipation_SW_linear_main}
\end{equation}
\end{small}
As for the AO and HO protocols, we report the minimum dissipation values (with respect to $\alpha$).

\begin{figure}[t!]
\includegraphics[width=1.1\linewidth]{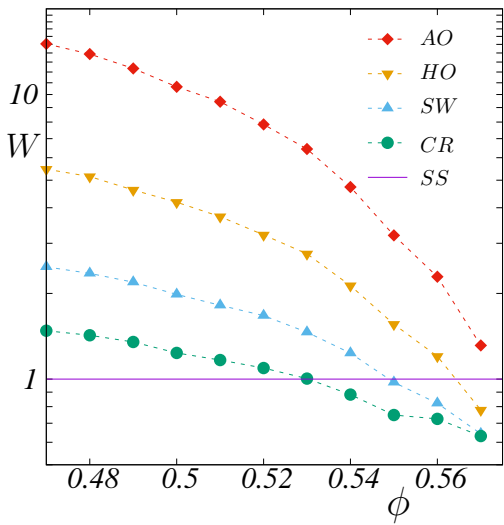}
\caption{Relative dissipation for different protocols as a function of packing fraction. We systematically decrease the dissipation by changing the protocol, most importantly by decreasing the accumulated strain in the perpendicular direction,
and conclude that cruising is better than the other protocols. Symbols correspond to numerical data using $\mu=1$, where viscosities from Fig.~\ref{fig_viscosity_allprotocol} have been used to find the corresponding minimum dissipation. Dashed lines are merely guides to the eye.
}
\label{fig_dissipation_allprotocols}
\end{figure}

\subsection{Cruising protocol}
Although the square wave protocol reduces the viscosity by a large amount (see section \ref{sec_results}), it will not propagate the suspensions when in cross mode, as the shearing along $xy$ and $-xy$ cancels any such propagation. The simplest way to overcome this issue is by continuously shear along one of these two opposite directions. This leads to 
 cruising protocol (CR), where, in the first half of the cycle, we shear the system along the flow direction of the $xz$ plane and in the second half along the flow direction of the $xy$ plane as shown in Fig. \ref{fig_sch}(d). 
 In contrast to all the above protocols, one main advantage of this protocol is that we propagate the suspension for each direction. 
In a full cycle, the ratio between the accumulated strain and the strain in the ``propagating'' direction equals $\sqrt{2}$ (compared to $\geq 2$ in the other protocols). Although the viscosity reduction here is the lowest among all the protocols, because of the highly efficient propagation, this is the most beneficiary in terms of dissipation per unit strain (and at a fixed global shear rate). We discuss the results later in the section \ref{sec_results} and make it clear that this protocol is best among all the protocols we discussed and equivalent to the best existing protocol called SO \cite{ness2018shaken}. The dissipation is related to the viscosities as  
\begin{small}
\begin{equation}
    W_{\rm CR} =\frac{1}{2} \left[\frac{\eta'}{\eta_{\rm ss}}\left(\frac{1}{(1-\alpha)}\right)+\frac{\eta''}{\eta_{ \rm ss}}\left(\frac{1}{\alpha}\right)\right].
    \label{eq_dissipation_CR_linear_main}
\end{equation}
\end{small}
In the above equation, we have relabeled the viscosities as $\eta'$ and $\eta''$ as none of them is now parallel or perpendicular to the propagating axis. Although $W_{\rm CR}$ looks like $2$ times smaller than $W_{\rm SW}$, it is important to understand that the viscosities depend on the protocol. 
Since we have $\eta'=\eta''=\eta$ in this protocol, the minimum dissipation occurs at $\alpha=0.5$. The minimum dissipation hence takes the simple form: $W_{\rm CR}=2\eta/\eta_{\rm ss}$. This generally means that we need at least a $50\%$ reduction of the viscosity in order to be a viable approach. 

\subsection{Relaxing the turn angle}
For all protocols, we have so far exclusively used $\pm 90^o$ turns with respect to the $xz$ plane when changing the flow direction. 
One can easily relax this constraint and allow for any degree of turns, with $\pm (1/2) \theta$ degrees of turns around the propagating direction. 
The dissipation for the CR protocol and using a particular angle $\theta$ can be written as (see the Appendix for details)
\begin{small}
\begin{equation}
   W_{\rm CR}(\theta) =\frac{1}{(2+2\cos\theta)} \left[\frac{\eta'}{\eta_{\rm ss}}\left(\frac{1}{1-\alpha}\right)+\frac{\eta''}{\eta_{\rm ss}}\left(\frac{1}{\alpha}\right)\right]. 
   \label{eq_dissipation_CR_angle_main}
\end{equation}
\end{small}
Note that it will yield Eq. (\ref{eq_dissipation_CR_linear_main}) at an angle $\theta = 90^0$. The reduced form, using $\alpha = 0.5$ and where $\eta'=\eta''=\eta$, is then
\begin{equation}
  W_{\rm CR}(\theta) = \frac{2}{(1+\cos\theta)} \frac{\eta}{\eta_{\rm ss}}  
 \end{equation}

which yields $W_{\rm CR}=1$ at $\theta=0^o$, as $\eta=\eta_{\rm ss}$ in that case. This perfectly agrees with the idea that cruising has now become a steady-shear protocol.

\section{ Simulation details}
\label{section_simulation}

\color{black}
The only unknowns in the above dissipation equations are the viscosities, which can be obtained by numerical simulations.
We simulate a dense athermal suspension with various protocols and measure the energy dissipation per strain and time. The latter is equivalent to having a fixed global shear rate in the transport direction. In this work, we study a non-inertial suspension composed of bi-disperse particles. The particles are iso-dense with the fluid and confined in a three-dimensional periodic box using Lees-Edwards boundary conditions at a packing fraction $\phi$.  We use a bi-disperse $50:50$ mixture of particles with a diameter ratio of $1:1.4$. A particle undergoes three types of force and torque: Stokes drag, contact, and lubrication forces. The Stokes drag force and torque are given by,
\begin{eqnarray}
\nonumber
 \vec{F}_i^s &=& -6\pi\eta_f a_i (\vec{v}_i-\vec{v}_f),\\
\vec{\tau}^s_i &=& -8\pi \eta_f a_i^3 (\vec{\omega}_i - \vec{\omega}_f).
\end{eqnarray}
 Here, $a_i$, $v_i$, and $\omega_i$ are the radius, linear velocity, and angular velocity of particle $i$, while $v_f$ and $\omega_f$ are the fluid's linear and angular velocity (vorticity) and $\eta_f$ is the viscosity of the fluid. We chose parameters such that the Stokes number $\rho \dot \gamma \langle a\rangle^2/\eta_f\ll1$ for $\alpha$ values in the range [0.2,0.8], where $\langle a\rangle$ denotes the average radius and $\rho$ the density of both solid particles and the fluid.
 
 The contact force between particles is described by harmonic springs
 \begin{equation}
\vec{F}_{ij}=k_n\delta^{ij}_n \hat{n}^{ij}+k_t\delta^{ij}_t \hat{t}^{ij}.
\end{equation}        
Here $\delta^{ij}_n$ is the normal overlap between two particles while $\delta^{ij}_t$ is the relative tangential displacement. The springs' normal and tangential stiffness constants are denoted by $k_n$ and $k_t$.
We choose $k_t=(2/7)k_n$, and $k_n$ and $k_t$ values large enough such that our particles full-fill near-hard sphere conditions (typically $k_n/P \sim 10^4$).
The unit vector along the normal and the tangential direction of two particles are denoted by $\hat{n}^{ij}$ and $\hat{t}^{ij}$. The tangential forces for each contact are restricted by the Coulomb friction criteria $|\vec{F}^t_{ij}|<|\mu\vec{F}^n_{ij}|$, where, $\mu$ is the coefficient of friction. When a lubricating film separates two particles, the interaction force acting along normal and tangential directions can be written as
\begin{eqnarray}
\nonumber
F_{ij}^{l,n}(h_{ij})&=&\frac{3}{2}\pi \eta_f a_{ij} \frac{(\vec{v}_i-\vec{v}_j)\cdot \hat{n}_{ij}}{h_{ij}+\delta},\\
F_{ij}^{l,t}(h_{ij})&=&\frac{1}{2}\pi \eta_f \ln\left(\frac{2 a_{ij} }{h_{ij}+\delta}\right)(\vec{v}_j-\vec{v}_i)\cdot \hat{t}_{ij},
\end{eqnarray}
where $a_{ij}$ is the effective radius defined as $a_{ij}=a_i a_j/(a_i+a_j)$ while $h_{ij}$ is the gap between two particles. We choose a lower cut-off $\delta=0.001 a_s$ (where $a_s$ is the radius of the smaller particle) to avoid divergence at $h_{ij}=0$. In our simulations, first, we create a packing with no overlap between particles. After applying various types of shear on such a system, we perform basic molecular dynamic simulations and solve the trajectories using LAMMPS \cite{plimpton1995fast}. All quantities are for steady-state conditions, where we sheared the suspension for at least  $2$ unit of accumulated strain before we start to measure. Data is then collected for a further $18$ unit of accumulated strains. Dissipation in various directions are measured by $W'=\int \mathbf{\sigma} : \mathbf{\dot \gamma}\, dt$. Unless specified, we strain each direction $1\%$ before we change direction (the exception being the AO, where we have an intermediate straining of $2\%$ for the intermediate part of the oscillation).
 \section{Results}
 \label{sec_results}
In this section, we discuss our numerical results. We focus on both the viscosity reduction as well as the dissipation for tested protocols.
 \subsection{Viscosity Reduction}
We begin by analyzing the viscosity reduction for all the protocols. Fig.~\ref{fig_viscosity_allprotocol}(a) and (b) show how the viscosities, in parallel and perpendicular directions, vary with the packing fraction for our studied protocols respectively. In the figure, we also plot the steady shear values of the viscosity for comparison. The figure shows that all
protocols successfully reduce the viscosity at a fixed packing fraction. The largest reduction is found for the AO scheme, while the least efficient turns out to be the case of the cruising. While the AO and CR protocols have roughly the same viscosities in both directions, as assumed in Ref.~\cite{ness2018shaken}, 
this is not a universal result, as both HO and SW protocols clearly show a lower perpendicular viscosity compared to their parallel ones.  
In general, the more reversals (in the perpendicular direction) we do between the primary strains, the larger the viscosity reduction.
This can be seen as AO corresponds to two reversals, HO to one, SW a half, and CR to zero. This agrees with Ref.~\cite{ness2018shaken}'s findings. \\

Interestingly, the SW and CR show very similar viscosities in the ``parallel'' direction but different in the ``perpendicular'' one \footnote{The notation of parallel and perpendicular in the cruising protocol is a bit misleading, as none of them is along the propagating axis. Nevertheless, treating them on the same footing is useful.}. Both protocols have the same accumulated strain in the perpendicular direction. The only difference is that the SW involves alternating directions (a half-shear reversal per cycle) along the perpendicular direction. Hence, a $1\%$ strain in the primary direction is not sufficient to lose the micro-structural memory built up during cross shears, explaining why we see a difference between these SW and CR viscosities. 

\begin{figure}[t]
\centering
\hspace*{-0.5cm}
\includegraphics[width=1.1\linewidth]{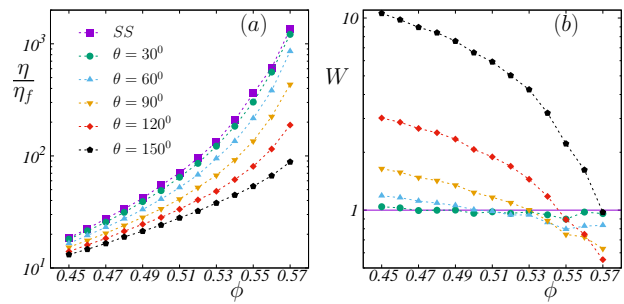}
\caption{ (a) Relative viscosity as a function of packing fraction for the cruising protocol at different cruising angles. It is clear that the viscosity reduction is larger for a wider angle, and it will hit a maximum at $180^o$. (b) Relative dissipation as a function of packing fraction in the cruising protocol for different cruising angles. The figure clearly shows that the angles at which the minimum dissipation occurs are different for different packing fractions. All the data are from direct simulation, using $\mu=1$ and $\alpha=0.5$.
}
\label{fig_dissipation_angle}
\end{figure}
\begin{figure}[t!]
\includegraphics[width=1.1\linewidth]{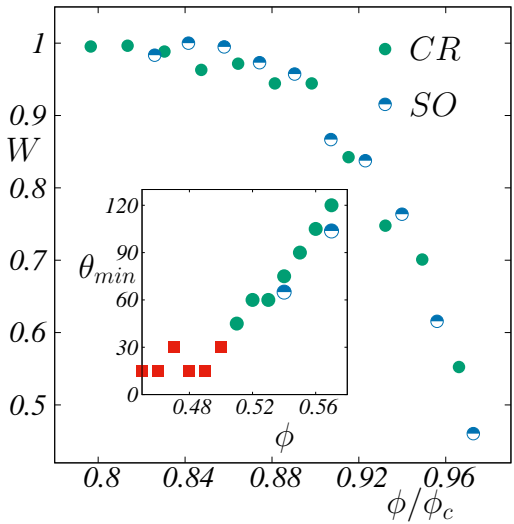}
\hspace{-0.2 in}
\caption{Comparison between the relative dissipation of SO and CR as a function of re-scaled packing fraction. The collapse indicates that both protocols are very similar to each other, and therefore ''random organization'' is not necessary to minimize the dissipation. {\bf (Inset)} The angle of minimum dissipation for different packing fractions. For the lowest $\phi$'s the relative dissipation is $1$ (within the noise) and marked as red square}
\label{fig_dissipation_compare}
\end{figure}
\subsection{ Dissipation reduction}
Suspensions dissipate energy when they flow, and the motivation of this work is to reduce the dissipation per unit strain at a fixed global shear rate. 
We here study the relative dissipation, \emph{i.e.~}we divide our dissipation with the corresponding steady-shear value at the same strain and global shear rate.
A relative dissipation below one indicates an energy saving when transporting the suspension. Fig.~\ref{fig_dissipation_allprotocols} shows the relative dissipation as a function of packing fraction for our tested protocols. Non-intuitively the viscosity and dissipation reduction are, in our cases, inversely correlated. The AO (corresponding to a full oscillation), which gives more than $90\%$ reduction in viscosity, turns out to be less energy efficient than steady simple shear. This is at odds with the original results of Ref.~\cite{ness2018shaken}, but it turned out that there were some subtle mistakes in that paper when calculating the dissipation for the AO protocol. The inefficiency of the AO comes from the fact that an extra oscillation is applied, which costs a huge amount of energy without propagating the suspension.
Only when we apply the CR or SW protocols we get a substantial energy saving for an extended range of packing fractions. Furthermore, compared to the SO protocol of Ref.~\cite{ness2018shaken}, another protocol that gives energy savings, we see that both SO and CR are superior in terms of energy saving, even if we for AO, HO, and SW have chosen the $\alpha$ giving the minimum dissipation. \\
\subsection{Effect of cruising angle}
We focus now on the CR protocol and relax the constraint with $90^o$ angles and test a few different angles with a scanning gap of $15^o$ \emph{i.e.} $\theta = 15^o, 30^o, 45^o....$ \emph{etc.}.
For example, $\theta=15^o$ means for the first half of a cycle, we shear the suspension along the flow direction of the $xz$ plane, while for the second half, we shear the system at an angle $15^o$ with respect to the $xz$ plane. 
Fig. \ref{fig_dissipation_angle}(a) shows the effect on viscosity, where the viscosity decreases monotonically as the angle widens, reaching its minimum value at $\theta =180^o$. However, as seen before, this information might
sometimes be misleading when analyzing dissipation. Fig~\ref{fig_dissipation_angle}(b) shows the relative dissipation as a function of packing fraction for five different angles. The tendency between the angles is clearly non-monotonic. For the lowest angle, the dissipation line is fairly flat and close to one, even though it is in general a few percent below one at intermediate to high packing fractions. At $\theta=0^o$, we have $W=1$, which is consistent with this protocol's physical and mathematical understanding as it is the same as the steady shear.   
Close to the shear jamming point for steady state, the dissipation seems to be lowest for angles close to $120-150^o$. The optimum angle varies with packing fraction, where a higher packing fraction favors wider angles and low packing fractions narrower angles. For very wide angles, the propagation becomes inefficient due to geometrical reasons, and the relative dissipation per strain diverges as $\theta\to180^o$. This limit corresponds to a pure oscillatory shear. In relative dissipation, the cruising protocol with an optimized angle is very similar to SO, and most likely within the numerical precision, for a wide range of packing fractions as shown in Fig. \ref{fig_dissipation_compare}. In the figure, we used $\phi/\phi_c$, for a better collapse, where $\phi_c$ is the packing fraction at which shear jamming occurs for the steady state. This since the $\phi_c$'s differ, due to microscopic details \cite{torquato2000random,Ness2022review} \footnote{We used two different $\phi_c$ values for our data and the data of Ref.~\cite{ness2018shaken}. Both $\phi_c$ values were extracted by fitting the steady state curves in with $\eta=\eta_0(\phi-\phi_C)^{-2}$. Our data yielded $\phi_c=0.5900$ while the data from Ref.~\cite{ness2018shaken} gave $\phi_c=0.5827$.}

\section{Discussion}
In this work, we have focused on the flow of dense suspensions and their dissipation. 
We found a new highly efficient shear cruising protocol that keeps the energy dissipation at a minimum.
We found that by adapting the cruising protocol's angle, we can get almost a magnitude in energy gain.
\\

Obviously, we need to comment on the success of the SO protocol. Re-analyzing that protocol, one sees that the SO protocol's minimum dissipation well corresponds to a CR motion with an optimum angle. The fact that the optimum angle varies with packing fraction (see inset of Fig.~\ref{fig_dissipation_compare}) corresponds well with the data of Ref.~\cite{ness2018shaken}, where the minimum in units of oscillation ratio $\omega \gamma/\dot \gamma$ (their units) varies with packing fraction.  Our re-interpretation is that this oscillation ratio can be translated into a cruising angle which we also show for two available packing fractions from Ref.~\cite{ness2018shaken} along with our minimum angles in the inset of Fig.~\ref{fig_dissipation_compare}. \\

Our results, hence, strongly suggest that the large dissipation reduction is not per see due to the oscillations, but rather due to a sudden change in direction.
Since we are dealing with frictional particles, a $>90^o$ turn will quite successfully deactivate frictional contacts. This picture challenges the idea that a random organization \cite{pine2005reverse,Corte2008random,Weijs2015revers} is the primary cause of dissipation reduction. Instead, our data suggest that dissipation reduction (or equivalent ``dissipative'' viscosity) is linked to dense suspensions' fragility and how it varies with both packing fraction, shear angles, and strains \cite{Cates1998fragile,seto19fragile,blanc23fragile}. Haven only considered a few protocols; this clearly raises the question of whether even more efficient protocols exist. We leave that as an open challenge for the
dense suspension community. \\
\newline
 
{\it Acknowledgments}: We thank Christopher Ness, Romain Mari, Joakim Stenhammar, and Zakiyeh Yousefian for the useful discussions. We also thank Christopher Ness for the open-access LAMMPS script. This project was funded by the Wenner-Gren Foundation (grant number UPD2021-0147). MT is financed by the Swedish Research Council (grant number 2021-04997).
The simulations were performed on resources provided by (storage) the Swedish National Infrastructure for Computing (SNIC) and the center for scientific and (computing) technical computing at Lund University (LUNARC).



\setcounter{section}{0}
\makeatletter 
\renewcommand{\thesubsection}{\@arabic\c@subsection}
\makeatother


\setcounter{equation}{0}
\makeatletter 
\renewcommand{\theequation}{A\@arabic\c@equation}
\makeatother

\section*{APPENDIX A: THEORETICAL CALCULATION\\ OF THE DISSIPATION}
\begin{figure}[h]
\centering
\hspace*{-0.5cm}
\includegraphics[width=1.1\linewidth]{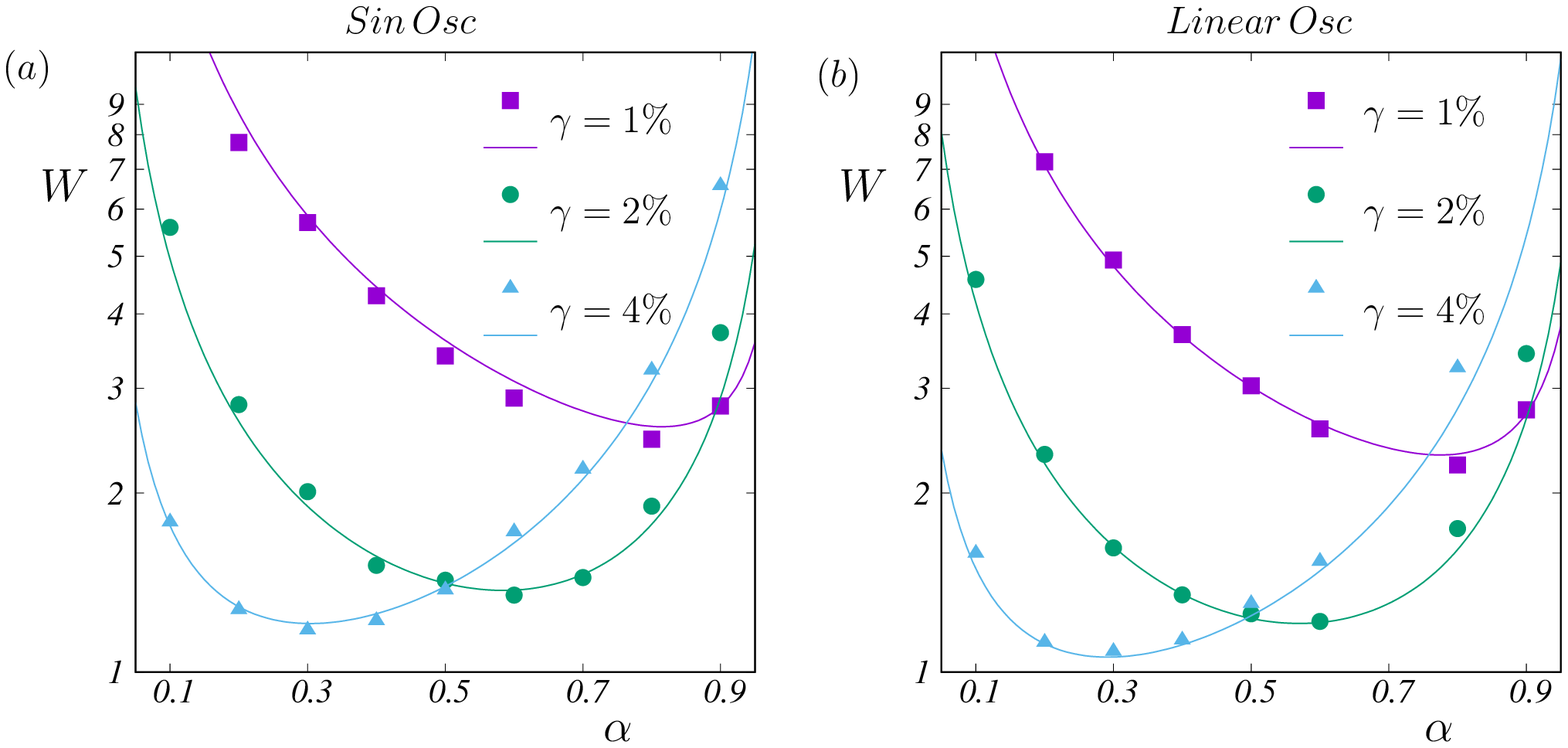}
\caption{
Relative dissipation for AO protocol with (a) sinusoidal oscillation along the perpendicular direction (b) linear oscillation along the perpendicular direction with various primary strains $\gamma_{\rm pri}$ and $\gamma_{\rm osc}=1\%$ fixed. Symbols are numerical data using $\mu=1$ and $\phi=0.56$. Lines are given by Eqs.~(\ref{eq_dissipation_AO_sinusoidal_appendix}) and (\ref{eq_dissipation_AO_linear_appendix}). Note that none of the tested strains' relative dissipation goes below one at least up to $\phi=0.56$.
}
\label{fig_dissipation_AO_appendix}
\end{figure}

\subsection{The AO protocol}

The dissipation per strain can be written as 
\begin{equation}
W_{\rm Tot} = W_{\parallel}+W_{\perp}
\end{equation} 
When comparing with the steady-state value, the relative dissipation is,
\begin{equation}
    W=\frac{W_{\rm Tot}}{W_{SS}} = \frac{W_{\parallel}+W_{\perp}}{W_{SS}}
    \label{eq_diss_appendix}
\end{equation}
The dissipation along the primary direction in a single cycle can be written as
\begin{equation}
W^{}_\parallel(per\,cycle) = \sum_0^{(1-\alpha)T} \eta_\parallel \dot{\gamma}^2_{\rm pri}, 
\label{eq_parallel_dissipation_cycle_appendix}
\end{equation}
where $\dot{\gamma}_{\rm pri}=\dot{\gamma}/(1-\alpha)$. Putting this in Eq.~(\ref{eq_parallel_dissipation_cycle_appendix}), we arrive at
\begin{equation}
W^{}_\parallel(per\,cycle) = \sum_0^{(1-\alpha)T} \frac{\eta_\parallel \dot{\gamma}^2}{(1-\alpha)^2} = \frac{\eta_\parallel \dot{\gamma}^2}{(1-\alpha)} T.
\end{equation}
Therefore, the dissipation per unit strain can be written as,
\begin{equation}
W_\parallel (per\,strain) = \frac{\frac{\eta_\parallel \dot{\gamma}^2}{(1-\alpha)} T}{\dot{\gamma}T} = \frac{\eta_\parallel \dot{\gamma}}{(1-\alpha)}.
\end{equation}
The viscosity along the perpendicular direction can be defined as

\begin{equation}
    \eta_\perp = \frac{\int_0^{2\pi/\omega}\sigma(t)\cos(\omega t) dt}{{\gamma}_{\rm osc} \omega\int_0^{2\pi/\omega}\cos^2(\omega t) dt} = \frac{W_{\rm OSC}(per\,cycle)}{(\gamma_{\rm osc} \omega)^2\int_0^{2\pi/\omega}\cos^2(\omega t) dt}.
\end{equation}

Therefore, the perpendicular dissipation in a full cycle takes the form,

\begin{equation}
    W_{\rm OSC}(per\,cycle)= \eta_\perp (\gamma_{\rm osc} \omega)^2\frac{\pi}{\omega}=\eta_\perp \gamma^2_ {\rm osc}\pi\omega.
\end{equation}
Putting $\gamma_{\rm osc}=\gamma_{\rm pri}=\dot{\gamma}T$ (1\% in our case) we arrive at
\begin{equation}
    W_{\rm OSC}(per\,cycle)= \eta_\perp \dot{\gamma}^2 T\frac{2 \pi^2}{\alpha}.
\end{equation}
Therefore dissipation per strain can be written as,
\begin{equation}
    W_{\perp}= \frac{2 \pi^2 \eta_\perp \dot{\gamma}}{\alpha}.
\end{equation}
The steady-state dissipation per strain is $W_{SS}=\eta_{ss} \dot{\gamma}$. Rewriting the Eq.~(\ref{eq_diss_appendix}) by putting $W_\parallel$, $W_\perp$, $W_{SS}$ we arrive at the final expression,
\begin{equation}
W = \frac{W_{\parallel}+W_{\perp}}{W_{\rm ss}}=\frac{\eta_\parallel}{\eta_{\rm ss}}\left(\frac{1}{1-\alpha}\right)+\frac{\eta_\perp}{\eta_{ss}}\frac{2\pi^2}{f^2}\left(\frac{1}{\alpha}\right),
\label{eq_dissipation_AO_sinusoidal_appendix}
\end{equation} 
which is Eq.~(\ref{eq_dissipation_AO_sin_main}) in the main text.

When a linear oscillation is applied along the perpendicular direction, the strain rate along the perpendicular direction needs to maintain such that in $\alpha T$ time, it covers a total of 4\% strain (1\% up, 2\% down and 1\% up again). Therefore, the strain rate along the perpendicular direction is related to the global strain rate as $\dot{\gamma}_{\rm osc}=4\dot{\gamma}/\alpha$. The dissipation along the perpendicular direction can then be written as
\begin{equation}
    W_{\rm OSC}(per\,cycle)= \sum_0^{\alpha T}\eta_\perp \dot{\gamma}_{\rm osc}^2 =  T\frac{16 \eta_\perp \dot{\gamma}^2}{\alpha}.
\end{equation}
Therefore, the dissipation per strain along the perpendicular direction takes the form
\begin{equation}
    W_{\perp}=   \frac{16 \eta_\perp \dot{\gamma}}{\alpha}.
\end{equation}
Putting this back in Eq.~(\ref{eq_diss_appendix}) yields
\begin{equation}
W_{\rm AO} = \frac{W_{\parallel}+W_{\perp}}{W_{\rm ss}}=\frac{\eta_\parallel}{\eta_{\rm ss}}\left(\frac{1}{1-\alpha}\right)+\frac{\eta_\perp}{\eta_{\rm ss}}\frac{16}{f^2}\left(\frac{1}{\alpha}\right),
\label{eq_dissipation_AO_linear_appendix}
\end{equation} 
which is Eq.~(\ref{eq_dissipation_AO_linear_main}) in the main text. We plot the relative dissipation as a function of $\alpha$ for both sinusoidal oscillation and linear oscillation along the perpendicular direction in Fig. \ref{fig_dissipation_AO_appendix}. The $\gamma$ in the figure legend is the strain in a single cycle along the primary direction, while we fix the maximum $\gamma_{osc}=1\%$ for all the curves. Therefore, $\gamma$ basically represents $f$ as discussed in the main text. A key thing to note from the figure is the dissipation is a little bit lower in the case of linear oscillation, as discussed in the main text and can be visualized from Eqs.~(\ref{eq_dissipation_AO_sinusoidal_appendix}) and (\ref{eq_dissipation_AO_linear_appendix}). It should be noted that although for different $f$, the perpendicular viscosity is similar, we used the exact values from the individual simulations. 
\subsection{The CR and SW protocols}
In the cruising protocol, the suspension is sheared along the flow direction of the $xz$ plane for a time period of $(1-\alpha)T$, which yields the same expression for the primary shear as above. As described in the main text, we relabeled the viscosities as $\eta'$ and $\eta''$ as none of them is now parallel or perpendicular to the propagating axis. Therefore, dissipation for the shearing along the primary direction in a cycle can be written as
\begin{equation}
W_{\rm PRI} (per\,cycle) =  \frac{\eta' \dot{\gamma}^2 T}{1-\alpha}.
\end{equation}
In a similar way, the dissipation per cycle for shearing at an angle $\theta$ can be written as
\begin{equation}
W_{\rm SEC} (per\,cycle) =  \frac{\eta''\dot{\gamma}^2 T}{\alpha}.
\end{equation}
However, unlike AO, the strain covered during a full cycle is now a vector sum of total strain in both directions and can be written as $\gamma=\sqrt{(2+2\cos(\theta))}\dot{\gamma}T$. The total dissipation per strain for both components can then be written as  
\begin{equation}
    W=\frac{1}{\sqrt{(2+2\cos(\theta))}}\left(\eta'\frac{\dot{\gamma}}{1-\alpha}+\eta''\frac{\dot{\gamma}}{\alpha}\right).
\end{equation}
As the global shear rate is changed, we need to divide this by the steady state value computed by the global shear rate. Therefore $W_{SS}$ per strain can be written as $W_{SS}=\sqrt{(2+2\cos(\theta))}\dot{\gamma}\eta_{ss}$. The final form of dissipation in the case of cruising is then 
\begin{equation}
   W_{\rm CR}(\theta) =\frac{1}{(2+2\cos\theta)} \left[\frac{\eta'}{\eta_{\rm ss}}\left(\frac{1}{1-\alpha}\right)+\frac{\eta''}{\eta_{\rm ss}}\left(\frac{1}{\alpha}\right)\right],
\end{equation}
which is Eq.~(\ref{eq_dissipation_CR_angle_main}) in the main text.
In the square wave protocol, we need to consider two cycles and take an average as the second half of the first cycle is different from the second half of the second cycle. In the second half of the first cycle, we shear the system along the flow direction of the $xy$ plane, while in the second cycle, we shear the system along the flow direction of the $-xy$ plane in the second half. This cancels the progression for the second half when we consider two complete cycles. Thus the strain it covers in two cycles along the flow direction of $xz$ plane is $2\dot{\gamma}T$, \emph{i.e.} in a single cycle it covers a $\dot{\gamma}T$ strain. Therefore, the global strain rate is also $\dot{\gamma}$, and by a similar calculation, we end up with the expression, 
\begin{equation}
    W_{\rm SW} = \frac{\eta_\parallel}{\eta_{\rm ss}}\left(\frac{1}{1-\alpha}\right)+\frac{\eta_\perp}{\eta_{\rm ss}}\left(\frac{1}{\alpha}\right),
\end{equation}
which is Eq.~(\ref{eq_dissipation_SW_linear_main}) in the main text.

\subsection{The SO protocol}
In the SO, unlike all the other protocols, the suspension is sheared in a primary direction and simultaneously a sinusoidal shear is applied perpendicular to this direction. In this case, the primary dissipation per strain can be written as
\begin{equation}
    W_{PRI} (per\,strain) = \eta_\parallel \dot{\gamma}.
\end{equation}
The perpendicular dissipation can be written as
\begin{equation}
    W_{\rm OSC} (per\,strain)= \frac{\eta_\perp \gamma_{\rm osc}^2\omega^2}{2\dot{\gamma}} 
\end{equation}
Adding both these dissipations and dividing by the steady-state dissipation per strain we finally arrive at
\begin{equation}
    W = \frac{\eta_\parallel}{\eta_{ss}} + \frac{\eta_\perp}{\eta_{ss}}\left(\frac{\gamma_{\rm osc}\omega}{\dot{\gamma}}\right)^2.
\end{equation}

\bibliography{vis}

\clearpage

 

\end{document}